# ElasTraS: An Elastic Transactional Data Store in the Cloud

*Sudipto Das    Divyakant Agrawal    Amr El Abbadi*
*Department of Computer Science, UC Santa Barbara, CA, USA*
{sudipto, agrawal, amr}@cs.ucsb.edu

## Abstract

Over the last couple of years, "Cloud Computing" or "Elastic Computing" has emerged as a compelling and successful paradigm for internet scale computing. One of the major contributing factors to this success is the *elasticity of resources*. In spite of the elasticity provided by the infrastructure and the scalable design of the applications, the elephant (or the underlying database), which drives most of these web-based applications, is not very elastic and scalable, and hence limits scalability. In this paper, we propose ElasTraS which addresses this issue of scalability and elasticity of the data store in a cloud computing environment to leverage from the elastic nature of the underlying infrastructure, while providing scalable transactional data access. This paper aims at providing the design of a system in progress, highlighting the major design choices, analyzing the different guarantees provided by the system, and identifying several important challenges for the research community striving for computing in the cloud.

## 1 Introduction

"Utility Computing" (popularly known in the industry as "Cloud Computing") has been an extremely successful model for providing *Infrastructure as a Service (IaaS)* over the internet, and has led to the tremendous success of companies such as Amazon as a technology provider through Amazon Web Services (AWS), Salesforce Inc. and many more. It has been widely discussed to be the "dream come true" for the IT industry, with the potential to transform and revolutionize the IT industry by making software even more attractive [2]. On one end of the spectrum are these *IaaS* providers that provide compute cycles, storage, network bandwidth etc., while on the other end of the spectrum are providers like Microsoft's Azure and Google's AppEngine who provide *Platform as a Service (PaaS)*, and are at a much higher level of abstraction. The term "Cloud Computing" encompasses this entire spectrum of services, but in this paper, we mainly concentrate on the *IaaS* models.

The major reasons for the widespread popularity and success of Cloud Computing are:

• *No up front cost and Pay-as-you-go model:* Allows new applications and product ideas to be tested easily and quickly without significant initial overhead.

• *Elasticity and illusion of infinite resources available on demand:* The *elastic* nature of the cloud allows resources to be allocated on demand allowing applications to easily scale up and down with load changes.

• *Transfer of risk:* Allows the handling of risk, e.g. failures, to be shifted from the smaller *Software as a Service* providers, to the larger entities, i.e. the cloud service providers, who are better equipped to mitigate the risks.

Typically, web-based applications have a 3-tier architecture, the *Web Server*, *Application Server*, and the *Database Server*. In general, different instances of *application servers* and *web servers* within the same application do not share any state information. Therefore, when the application load increases, the *application server* layer and the *web server layer* can be easily scaled up by spawning new machine instances that absorb the increased load. But in most common cases, the database back-end becomes the scalability bottleneck, since the database servers do not easily scale. In such a scenario, if the database server also had the elastic property of scaling up and down as per the load characteristics, then the entire software stack would scale better. One might argue in favor of data management services like Amazon's SimpleDB which can scale to huge amounts of data and large number of requests, but SimpleDB and similar scalable *key-value* stores like Bigtable [7] and Dynamo [10], although highly scalable, stop short of providing transactional guarantees even on a single row. In this paper, we propose ElasTraS, an *Elastic Transactional Data Store*, which is elastic along the same lines as the elastic cloud, while providing transactional guarantees. ElasTraS is designed to be a light-weight data store that supports only a sub-set of the operations supported by traditional database systems, and hence we call ElasTraS a *data store*, while reserving the term *databases* for more traditional database systems. ElasTraS is analogous to partitioned databases [3] which are common in enterprise systems, while adding features and components critical towards *elasticity* of the data store. In our design, we lever-age from proven database techniques [21, 12] for dealing with concurrency control, isolation, and recovery, while using design principles of scalable systems such as Bigtable [7] to overcome the limitations of distributed database systems [15, 18]. This paper aims at providing the design of a system in progress, highlighting the major design choices, analyzing the different guarantees provided by the system, and identifying new challenges for the research community.

## 2 Related Work

Based on the expertise gained from building distributed database systems [15, 18], researchers and designers have realized that supporting distributed transactions does not allow scalable and available designs. Hence, to satisfy the scalability requirements of web applications, designers have sacrificed the ability to support distributed transactions [13]. This resulted in the design of simpler data stores based on the *key-value* schema, where tables are viewed as a huge collection of *key-value* entries, and the values might have some structure, or may be viewed as uninterpreted strings of bytes [9]. Examples of these systems include Bigtable [7], Dynamo [10], PNUTS [8], Amazon SimpleDB, Facebook Cassandra, and many more. These systems limit access granularity to single key accesses, while providing minimal consistency and atomicity guarantees on multi-key accesses. This allows the system to horizontally partition the tables, without worrying about the need for distributed synchronization. These systems can be viewed to be the other end of the spectrum of data management solutions when compared to transactional databases. In particular, these systems are designed for high scalability and availability, but not to replace the traditional databases for consistent transactional properties, and recovery.

Other systems spanning the middle ground include Sinfonia [1], where the *minitransaction* primitive demonstrates that distributed transactions and distributed commitment, if used in a prudent manner, can be used for the design of scalable distributed systems. Similarly, Chubby [5], which uses the expensive Paxos consensus protocol [14], forms the core of scalable data management systems such as Google File System [11] and Bigtable [7]. These systems demonstrate that even though the paradigm of distributed databases was not successful, a lot of important concepts that were developed for these systems can be effectively used in modern scalable systems.

There have been efforts in designing databases for the cloud. Brantner et al [4] suggest a design for a database server that can use Amazon's S3 as a storage layer for the database engine, while the transaction manager might be located inside or outside the cloud. On the other hand, Lomet et al. [16] suggest a radically different approach where they suggest unbundling the transaction and the data manager. Even though these techniques open many interesting research avenues, the authors do not address the problem of elastic scaling of database systems in the cloud. ElasTraS has been designed with the goal of *scalable* and *elastic* transaction management in the cloud.

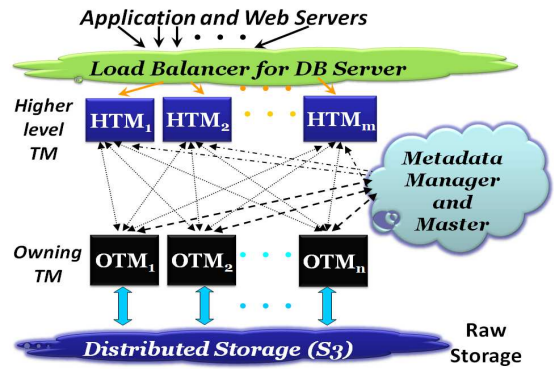

Figure 1: Overview of the ElasTraS system.

## 3 System Design

### 3.1 Data Model

Practice has shown that most application developers do not require a flexible schema as defined in the SQL standard and supported by most database systems. Modern application developers, in most cases, need systems with simple schema for storing their data, while providing fast and efficient access to it. The flexible data model supported by conventional databases is an overkill in most cases [10, 19]. Modern applications need a slightly enhanced version of *Indexed Sequential Access Methods*, and this forms the basis for the data model of modern scalable systems like Bigtable [7], Dynamo [10], PNUTS [8], SimpleDB etc. For ElasTraS, we choose a *key-value* based design similar to Bigtable [7] where *values* have application specified structure.

### 3.2 Design Overview

The ElasTraS system has been designed with the intent to provide transactional guarantees in a scalable manner, rather than retrofitting these features into an existing system. Figure 1 provides a high-level overview of the design of ElasTraS. At the heart of the system is a two-level hierarchy of *Transaction Managers (TM)* which are responsible for providing transactional guarantees, while providing elastic scalability with increase in demand. At the top of the stack are the application servers and the web servers that interact with the database. Requests to the database are handled through the *load balancer*. When a transaction request arrives, the load balancer forwards it to a *Higher Level Transaction Manager (HTM)* based on some load balancing policy. The HTM then decides whether it can execute the transaction locally, or route it to the appropriate *Owning Transaction Manager (OTM)* which *owns* exclusive access rights to the data accessed by the transaction. The actual data for the data store is stored in the distributed storage layer. All critical state information of the system, i.e. the *system state* [9] and the *metadata* for the tables, is managed by the *Metadata Manager*. All the components of ElasTraS are located in the cloud. In ElasTraS the database tables are partitioned, and ElasTraS can be configured for both *static* and *dynamic* partitioning. *Static partitioning* in ElasTraS is analogous to database partitioning [3] – the database designer partitions the database, and ElasTraS is responsible for mapping partitions to specific OTMs, and reassigning partitions with changing load characteristics to ensure scalability and elasticity. In this configuration, the application is aware of the partitions, and hence can be designed to limit transactions to single partitions. In such a configuration, ElasTraS can provide ACID transactional guarantees for transactions limited to a partition. Under *dynamic partitioning* configuration, ElasTraS, in addition to managing partition mapping, is also responsible for database partitioning using range or hash based partitioning schemes. Since the applications are not aware of the partitions, transactions are not guaranteed to be limited to a single partition. To ensure scalability in a *dynamic partitioning* setup, and avoid distributed transaction, ElasTraS only supports *minitransactions* [1], a scalable and primitive with restricted transactional semantics, which ensures recovery but not global synchronization. We now discuss the different components of the system, and Section 4.1 provides details of various transactional guarantees provided by ElasTraS.

### 3.2.1 Distributed Storage

The distributed storage layer provides an abstraction of a fault-tolerant shared disk which can be accessed from anywhere in the network. This abstraction of the storage layer has effectively been used in the design of a number of systems such as the use of the Google File System [11] by Bigtable [7] and the use of Amazon S3 in designing a database on S3 [4]. The storage layer takes care of replication and fault-tolerance, while the application accessing the storage should ensure that the same object or file is not being written to concurrently. Considering the fact that synchronous replication is expensive, it can be expected that the storage layer replication will be asynchronous and *eventually consistent* [20]. But if there are a limited number of failures, it can be easily assumed that the storage layer provides consistent access to single objects, or in other words, reads and writes to the storage layer are atomic. Since in the presence of failures, the storage layer might return stale data, some notion of versions should be associated with the data. If the storage layer provides support for versioning, then the system can leverage it, otherwise, versioning should be explicitly incorporated.

### 3.2.2 Owning Transaction Managers

The *Owning Transaction Managers (OTM)* are the entities responsible for the execution of transactions on the partitions of the databases, and have exclusive access rights to the partitions they own. These are analogous to the *tablet servers* in Bigtable [7], and *own* disjoint partitions of the database. An OTM is responsible for all the concurrency control and recovery functionality for the partitions it owns. Since an OTM has exclusive access to the set of partitions its *owns*, it can aggressively cache the contents of the partition in its local disk, thereby preventing expensive accesses to the distributed storage which actually stores the data. To guarantee the durability of committed transactions, all changes made by a transaction should be stored on some medium that can tolerate the failure of the OTM, and allow the system to recover from such failures and guarantee the durability of committed transactions.

In order to deal with dynamic partition assignments and the failure of OTMs, and to ensure that only one OTM is serving a partition, every OTM acquires a *lease* for a partition with the *metadata manager*, which is renewed periodically. If an OTM has successfully acquired a lease, then the *metadata manager* ensures that the OTM has exclusive access to the partition, and hence the OTM can execute transactions on the partition without the need for distributed synchronization. This mechanism of distributed lease maintenance is similar to that supported by Chubby [5] and used in Bigtable [7].

### 3.2.3 Metadata Manager and Master

The *Metadata Manager and Master (MMM)* is the brain of the system that stores the *system state* [9], viz., partition information, mapping of partitions to OTM, leasing information for the OTMs to deal with failures, and monitoring the health of the system. In addition to providing strong durability and consistency guarantees for the metadata of the system, this entity also acts as a *Master* which monitors the health of the system and performs the necessary system maintenance in the presence of failures. The *Master* monitors the system and ensures that if an OTM fails, then another OTM is instantiated to serve the partition, and also deals with partition reassignment for load balancing. High consistency of the data stored in the MMM is guaranteed through synchronous replication of the contents. For this purpose, we choose a system design similar to that of the Chubby locking service [5] that uses the Paxos consensus algorithm [14] for replica con-sistency [6]. Paxos guarantees safety and consistency in the presence of arbitrary failures, but the availability or liveness of the system is not guaranteed in the presence of failures. In the AWS infrastructure, failures in one *availability zone* are isolated from failures in other *availability zones*. Hence, high availability can be achieved by the judicious placement of replicas so that correlated failures do not affect a majority of the replicas. Note that the presence of Paxos [14] in the core makes write accesses to the MMM costly. As a result, the MMM should not be heavily loaded with a huge number of requests. Since the MMM does not reside in the data path, and in most cases the clients of the system can cache the metadata, the MMM should not be a performance bottleneck for the system.

#### 3.2.4 Higher level Transaction Managers

The *Higher level Transaction Managers (HTM)* are designed to absorb all read-only transactions in the workload. HTMs cache subsets of the database for read-only purposes, and answer queries from its cache. In *static partitioning*, transactions associated with a single partition are routed to the appropriate OTM which executes the transactions. For *minitransactions* [1], the HTM becomes the *coordinator*, while OTMs owning the partitions accessed by the *minitransaction* are the *cohorts*. Neither readonly transactions nor *minitransactions* associate any state with the HTMs. In both cases, the HTMs do not have any state coupling with OTMs, and the number of OTM and HTM instances can be different depending on the system configuration and the load on the system. For routing a request to the appropriate OTM, the HTM caches the mapping of partitions to OTMs. Unlike the OTMs, which are responsible for only partitions of the database, the HTMs span the entire database for read-only queries and *minitransactions*.

## 4 Implementation Sketch

### 4.1 Transaction Management

In a *statically partitioned* setup, applications can limit transactions to single partitions. Within a partition, ElasTraS provides ACID guarantees similar to transactions in databases. The only subtle difference is in the level of consistency (the C in ACID) guaranteed by ElasTraS. To obviate distributed synchronization, and minimize the impact of a single TM failure on the operation of the remaining TMs, ElasTraS guarantees consistency only within a partition of the database and there is no notion of consistency *across* partitions or global serializability [21]. Efficient performance can therefore be achieved since an OTM is guaranteed exclusive access to the partitions it *owns*, and proven techniques for concurrency control, isolation, and recovery in traditional databases [21, 12] can be used in designing the OTM.

Since an OTM has exclusive access to the partition, it can aggressively cache the database contents in its local disk, and the updates can be asynchronously applied to the distributed store in a manner similar to database checkpointing [21]. Since *dynamic partitioning* does not allow applications to limit transactions to a single partition, ElasTraS only supports *minitransactions* [1] in this setup, while supporting both forms of transactions for a *statically partitioned* setup. In executing a *minitransaction*, there is no need for synchronization amongst participating OTMs, and hence, *minitransactions* can be efficiently executed within ElasTraS.

### 4.2 Recovery done right

**Failure of OTM:** To ensure the durability of transactions, OTMs use write-ahead logging. As a result, techniques such as ARIES [17] and other similar recovery techniques [21, 12] can be used for efficient recovery. In the AWS model, when a machine instance crashes, its local disk is also lost. Therefore, to guarantee the durability and persistence of transactions beyond the failure of an OTM, the log entries of committed transactions should be stored on some persistent storage. But since logging must be performed during normal transaction execution, the performance of an OTM would suffer if the log entries are *forced* to a distributed store like S3. AWS provides a better solution in the form of Elastic Block Storage (EBS), which provides persistence beyond instance failures, allowing log entries to be stored on EBS. Our initial experiments with EBS showed that for sequential reads and writes, the performance of EBS is comparable to disks associated with AWS machine instances. Therefore, using EBS for both caching data pages, and storing logs would provide better performance, though at a higher dollar cost. When an OTM fails, its lease with the *metadata manager* expires, and once the *master* notices that the OTM instance has failed, it can instantiate a new OTM, which recovers the state of the failed OTM from the logs. A similar mechanism is used in Bigtable [7] to deal with *tablet server* failures.

**Failure of HTM:** Very little state is associated with the HTMs and hence, handling HTM failures is easier. All progress of a read-only transaction is lost when the HTM executing it fails. Clients of such transactions time out, and application specific recovery triggered. For *minitransactions* [1], no state is associated with the *coordinator*, and hence the recovery principles described in Sinfonia [1] can be used to recover a *minitransaction* from HTM or *coordinator* failures.

### 4.3 Elasticity

The ability of the system to deal with dynamic partition reassignments is the basis for the *elasticity* of the data store. When the load on the system is low, the system can operate with a small number of HTM and OTM instances. As the load on the system increases, the partitions can be reassigned to lesser loaded OTMs, or new OTMs can be spawned to absorb the load and take ownership of some partitions from the heavily loaded transaction managers. Since the HTMs do not share any state information, spawning new HTMs is easy, and amounts to spawning new instances that can immediately start absorbing the load. On the other hand, when a new OTM instance is created, it has to acquire a lease from the *metadata manager*, obtain control for some of the partitions, update the *metadata* to reflect the modified mapping, and perform any partition specific recovery. Once this is done, the new OTM is ready to execute transactions for the partition, while the old OTMs reject or redirect transactions of the reassigned partitions. Similarly, when the load decreases, the system can detect this and decide to remove OTM and/or HTM instances. Removal of HTM instances requires notifying the load balancer about the update, and then stopping the physical instance. Removing OTM instances requires updates to the *metadata* and the transfer of control of the partitions to some other OTM. Once control is transferred, the physical OTM instance can be stopped. Since AWS uses a *pay-as-you-go* model, stopped instances do not incur any cost, and hence, at low loads, the cost of operating the data store also reduces.

## 5 Discussion and Conclusion

The proposed design of ElasTraS can provide ACID guarantees for transactions that are limited to a single partition. Practice has shown that most web workloads are limited to single object accesses [19, 10], and hence, these transactional accesses can be trivially supported by ElasTraS. In addition, most enterprise systems are designed for statically partitioned databases, and transactions are limited to a single database partition, and ElasTraS can provide efficient, scalable, and elastic transactional access to the partitioned data store. In such a scenario, the ElasTraS design can still provide *elasticity* and *scalability* by dynamic partition reassignment based on the load on the system, while providing serializable transactional guarantees within a partition. Additionally, for applications where the designer does not want to statically partition the database, ElasTraS can dynamically partition the database. In addition to *single-object* transactions, ElasTraS can be easily extended to support *minitransactions* as defined by the Sinfonia system [1], but flexible transactions are not supported in the interest of scalability and elasticity. Based on these requirements of modern applications, we are in the process of formalizing the various forms of transactions that can be efficiently executed by ElasTraS, while having minimal impact of application design and preserving the 3-tier architecture of *web-servers*, *application servers*, with ElasTraS replacing the *database servers* in the cloud, thus providing a high degree of *elasticity* and *flexibility* throughout the entire architecture.

## Acknowledgements

The authors would like to thank the anonymous reviewers for their useful comments and suggestions that helped improve this paper. This work is partly supported by NSF Grants IIS-0744539 and CNS-0423336